\documentclass[manuscript]{aastex6}
\usepackage[utf8]{inputenc}
\usepackage{wasysym}
\usepackage{afterpage}
\usepackage{color}

\begin{document}

\title{DIFFERENCES IN THE GAS AND DUST DISTRIBUTION IN THE TRANSITIONAL DISK OF A SUN-LIKE YOUNG STAR, PDS 70}

\author{Zachary C. Long\altaffilmark{1}, Eiji Akiyama\altaffilmark{2}, Michael Sitko\altaffilmark{1,3}, Rachel B. Fernandes\altaffilmark{4}, Korash Assani\altaffilmark{1}, Carol A. Grady\altaffilmark{5}, Michel Cure\altaffilmark{6}, Ruobing Dong\altaffilmark{7,8}, Misato Fukagawa\altaffilmark{9}, Yasuhiro Hasegawa\altaffilmark{10}, Jun Hashimoto\altaffilmark{2}, Thomas Henning\altaffilmark{11}, Shu-Ichiro Inutsuka\altaffilmark{12}, Stefan Kraus\altaffilmark{13}, Jungmi Kwon\altaffilmark{14}, Carey M. Lisse\altaffilmark{15}, Hauyu Baobabu Liu\altaffilmark{16}, Satoshi Mayama\altaffilmark{17}, Takayuki Muto\altaffilmark{18}, Takao Nakagawa\altaffilmark{19}, Michihiro Takami\altaffilmark{20}, Motohide Tamura\altaffilmark{2,21,22},  Thayne Currie\altaffilmark{23}, John P. Wisniewski\altaffilmark{24}, Yi Yang\altaffilmark{2,25}}

\altaffiltext{1}{Department of Physics, University of Cincinnati, Cincinnati, OH 45221, USA}
\altaffiltext{2}{National Astronomical Observatory of Japan, 2-21-1, Osawa, Mitaka, Tokyo, 181-8588, Japan}
\altaffiltext{3}{Center for Extrasolar Planetary Studies, Space Science Institute,  4750 Walnut St, Suite 205, Boulder, CO 80301}
\altaffiltext{4}{Lunar and Planetary Laboratory, University of Arizona, Tucson, AZ 85719}
\altaffiltext{5}{Eureka Scientific, 2452 Delmer St. Suite 100, Oakland CA 96402, USA}
\altaffiltext{6}{Instituto de Física y Astronomía, Universidad de Valparaíso, Chile}
\altaffiltext{7}{Steward  Observatory,  University  of  Arizona,  Tucson,  AZ, 85719}
\altaffiltext{8}{Bok Fellow}
\altaffiltext{9}{Division of Particle and Astrophysical Science, Graduate School of Science, Nagoya University, Furo-cho, Chikusa-ku, Nagoya, Aich 464-8602, Japan}
\altaffiltext{10}{Jet Propulsion Laboratory, California Institute of Technology, Pasadena, CA 91109, USA}
\altaffiltext{11}{Max Planck Institute for Astronomy, Königstuhl 17, D-69117 Heidelberg, Germany}
\altaffiltext{12}{Department of Physics, Graduate School of Science, Nagoya University, Nagoya 464-8602, Japan}
\altaffiltext{13}{School of Physics, Astrophysics Group, University of Exeter, Stocker Road, Exeter EX4 4QL, UK}
\altaffiltext{14}{Institute of Space and Astronautical Science, Japan Aerospace Exploration Agency, 3-1-1 Yoshinodai, Chuo-ku, Sagamihara, Kanagawa 252-5210, Japan}
\altaffiltext{15}{Space Exploration Sector, JHU-APL, 11100 Johns Hopkins Road, Laurel, MD 20723, USA}
\altaffiltext{16}{European Southern Observatory (ESO), Karl-Schwarzschild-Str. 2, 85748 Garching, Germany}
\altaffiltext{17}{The Center for the Promotion of Integrated Sciences, The Graduate University for Advanced Studies~(SOKENDAI), Shonan International Village, Hayama-cho, Miura-gun, Kanagawa 240-0193, Japan}
\altaffiltext{18}{Division of Liberal Arts, Kogakuin University, 1-24-2 Nishi-Shinjuku, Shinjuku-ku, Tokyo, 163-8677, Japan}
\altaffiltext{19}{Institute of Space and Astronautical Science, Japan Aerospace Exploration Agency, Kanagawa 252-5210, Japan nakagawa@ir.isas.jaxa.jp}
\altaffiltext{20}{Institute of Astronomy and Astrophysics, Academia Sinica, P.O. Box 23-141, Taipei 10617, Taiwan}
\altaffiltext{21}{Department of Astronomy and RESCUE, The University of Tokyo, 7-3-1, Hongo, Bunkyo-ku, Tokyo, 113-0033, Japan}
\altaffiltext{22}{Astrobiology Center of NINS, 2-21-1, Osawa, Mitaka, Tokyo, 181-8588, Japan}
\altaffiltext{23}{National Astronomical Observatory of Japan, Subaru Telescope, National Institutes of Natural Sciences, Hilo, HI 96720, USA}
\altaffiltext{24}{Homer L. Dodge Department of Physics, University of Oklahoma, Norman, OK 73071, USA}
\altaffiltext{25}{Department of Astronomical Science, The Graduate University for Advanced Studies ~(SOKENDAI), 2-21-1 Osawa, Mitaka, Tokyo 181-8588, Japan}

\begin{abstract}

We present ALMA 0.87 mm continuum, HCO\textsuperscript{+} J=4--3 emission line, and CO J=3--2 emission line data of the disk of material around the young, Sun-like star PDS 70. These data reveal the existence of a possible two component transitional disk system with a radial dust gap of $0\farcs42 \pm 0\farcs05$, an azimuthal gap in the HCO\textsuperscript{+} J=4--3 moment zero map, as well as two bridge-like features in the gas data. Interestingly these features in the gas disk have no analogue in the dust disk making them of particular interest. We modeled the dust disk using the Monte Carlo radiative transfer code HOCHUNK3D \citep{whitney13} using a two disk components. We find that there is a radial gap that extends from 15-60 au in all grain sizes which differs from previous work. 
\end{abstract}

\section{Introduction}

Large interferometer telescopes such as the Atacama Large Millimeter/submillimeter Array (ALMA) and the Very Large Telescope Interferometer (VLTI) have recently facilitated an increasingly detailed examination of circumstellar disks around young stars. They have the ability to detect molecular emission spectra, as well as dust continuum emission, at very high angular resolutions even at the millimeter and sub-millimeter wavelengths. The much higher angular resolution, when compared to a single telescope at similar wavelengths, allows us to probe closer to the central star, and study the spatial relationships between gas and dust on a finer scale than ever before. 

PDS 70 is a pre-main-sequence star with spectral type K5 and assumed to be part of the Centaurus association, estimated to be at a distance of 140 pc \citep{riaud06}. It has previously been imaged in both $H$-band polarized light using Subaru \citep{hashimoto12}, as well as more recently at 1.3 mm with the Sub Millimeter Array (SMA) \citep{hashimoto15}. In this paper we present a detailed examination of the circumstellar disk of PDS 70 in the 0.87 mm continuum, CO J=3--2 emission, and HCO\textsuperscript{+} J=4--3 emission based on ALMA observations. These data are at a much higher resolution than the SMA data, $0\farcs19 \times 0\farcs15$ compared to $0\farcs88 \times 0\farcs43$, and provide us with a good probe of the larger grains within the disk. They also allow us to examine the gas distribution and detect fine structures within disk gas. This is of particular interest to our study due to the existence of features which appear in the gas disk but are not present in the dust disk. PDS 70 has proven to be a system with a different structure in each of our continuum, CO J=3--2,  HCO\textsuperscript{+} J=4--3 measurements presented below. In addition to our observational data we present a Monte Carlo Radiative Transfer (MCRT) model of the disk using HOCHUNK3D \citep{whitney13} to simultaneously fit existing imagery as well as the spectral energy distribution (SED) of PDS 70. Following the discussion of the data and model, we also present additional SpeX and REM observations of PDS 70 as well as analysis of these data sets in the Appendix including gas accretion, mass, and age estimates.

\section{Observations and Data Reduction}

Observations were conducted on August 14 and 18, 2016 in the ALMA cycle 3 program (No. 2015.1.000888.S). We used 38 to 39 of the 12 m diameter antennas in the observation. The array configuration provides a maximum and minimum baseline length of 1.5 km and 15.1 m, respectively. Sky conditions were relatively stable with the precipitable water vapor between 0.45 mm and 0.82 mm. The system temperature was between 90 K and 280 K during observations. The correlator was tuned at the center frequency of 345.761 GHz for CO J=3–-2, 356.698 GHz for HCO\textsuperscript{+} J=4–-3, and 343.955 GHz and 355.497 GHz for continuum observations in dual polarization mode. The effective bandwidths for both emission lines are 468.75 MHz and their frequency resolution is 244.141 kHz, corresponding to approximately 0.21 km s\textsuperscript{-1} in the velocity resolution.

Data reduction was performed using version 4.7 of the Common Astronomy Software Applications package (CASA). A quasar J1427-4206 was used for bandpass, flux, and phase calibrations, and another quasar J1407-4302 was selected as a second calibrator. We used the task CLEAN in CASA and a continuum image was generated by combining the line-free channels in all of the spectral windows. The self-calibration was performed by applying the obtained continuum image as a model, and gain calibration was repeated until the rms reached minimum. The gain table obtained after the self-calibration of the continuum data was applied to both the CO J=3--2 and HCO\textsuperscript{+} J=4--3 to generate self-calibrated visibilities. Because the line and continuum observations were simultaneously conducted, and the atmospheric variation equally affects the phase variation of line and continuum data, the calibration table obtained from the continuum data can be applied to line emission calibration. Natural weighting was applied in both continuum and line imaging for better sensitivity. The achieved rms was 0.0551 mJy beam$^{-1}$ for the continuum, 1.204 mJy beam$^{-1}$ for CO J=3--2, and 1.118 mJy beam$^{-1}$ for HCO\textsuperscript{+} J=4--3. The field of view was approximately 18$\arcsec$ and the final synthesized beam of approximately $0\farcs19\times0\farcs15$ for the continuum image and CO J=3--2 map, and $0\farcs18\times0\farcs15$ for the HCO\textsuperscript{+} J=4--3 map after calibration and flagging as denoted in Figures~\ref{fig:continuum},~\ref{fig:gas_data0}. The total integration time on source after flagging is about 56 minutes.

\section{Results}

\subsection{0.87 mm Continuum Measurements}

\begin{figure}[ht]
\centering
\includegraphics[scale=0.7]{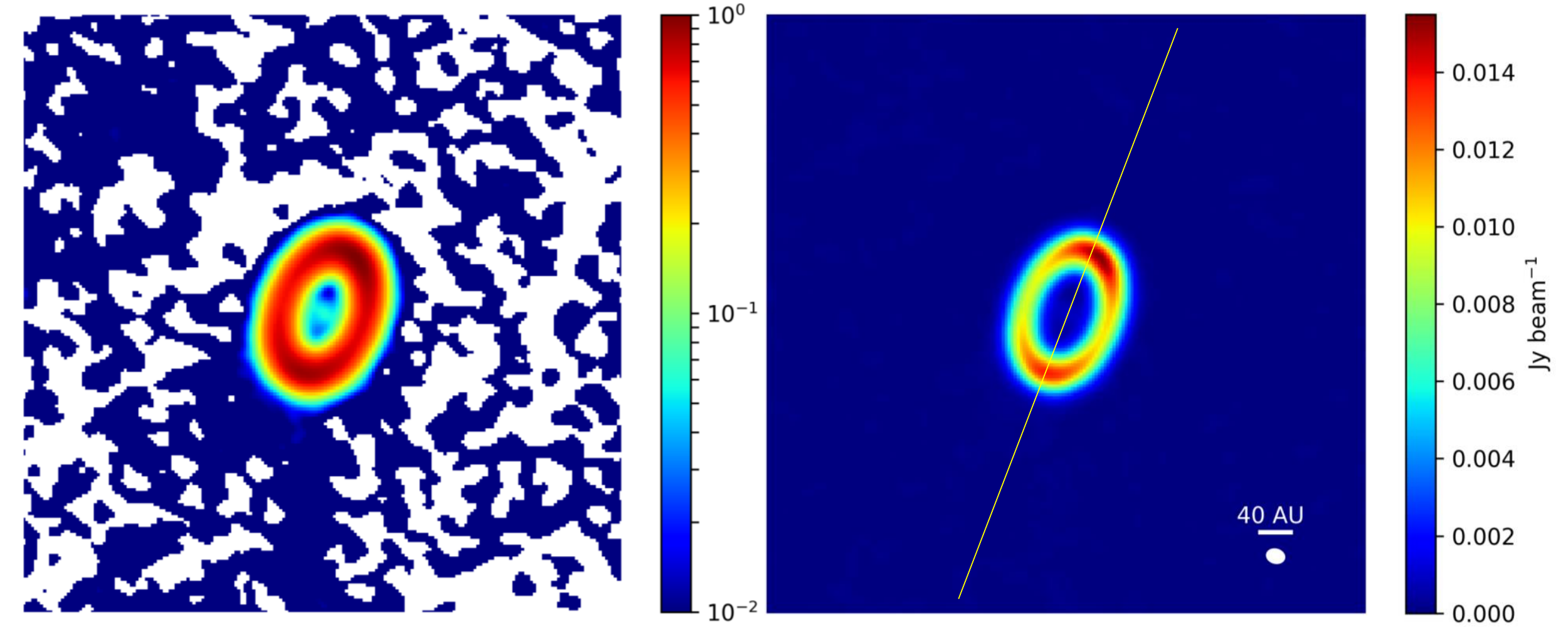}
\caption{Left: 0.87 mm continuum ALMA image with intensity normalized and log scaled to show inner disk component. Pixels with a scaled intensity less than 0.01 are removed to make the scale more useful. Right: 0.87 mm continuum image with a yellow line overlaid showing the major axis of the disk which lies at 159 $\pm$ 5$\degr$. North is up and East is left in these images.}
\label{fig:continuum}
\end{figure}

The 0.87 mm continuum image gives us a good measure of the radial extent of the dust disk, the flux density, and reveals some other interesting features. The major axis of the disk lies at $\sim$159$\degr$ East of North, measured via ellipse fitting of the brightest emission along the ring of the disk, (Figure~\ref{fig:continuum}) and measures a radial distance of $0\farcs42 \pm 0\farcs05$ ($\sim$60 au) from the center at the edge and $0\farcs79 \pm 0\farcs05$ ($\sim$110 au) at the outer edge. Both edges were found by measuring a distance equal to half the width of the beam size from the inner and outer edges of the disk that are detectable above noise. The measured flux density is 233 $\pm$ 3 mJy integrated at a radius of $1\farcs00$, past which there is no detectable emission above $\sim 2 \sigma$. In Figure~\ref{fig:continuum} the disk brightness is clearly not symmetrical about the minor axis, similar to what is seen in \citet{kraus17}, with the NW side $\sim$15$\%$ brighter than the SE side. 

There also appears to be material visible near the center of the disk in the continuum, seen more clearly in the log scaled image (Figure~\ref{fig:continuum},  left), which is separated from the outer disk by a radial gap. It is slightly larger than the resolution of the measurement, extending out approximately 0.2$\arcsec$ where the resolution of the instrument is $0\farcs19 \times 0\farcs15$ and is likely an inner disk component.

\subsubsection{HCO\textsuperscript{+} J=4--3 Measurements}

\begin{figure}[ht]
\centering
\includegraphics[scale=0.45]{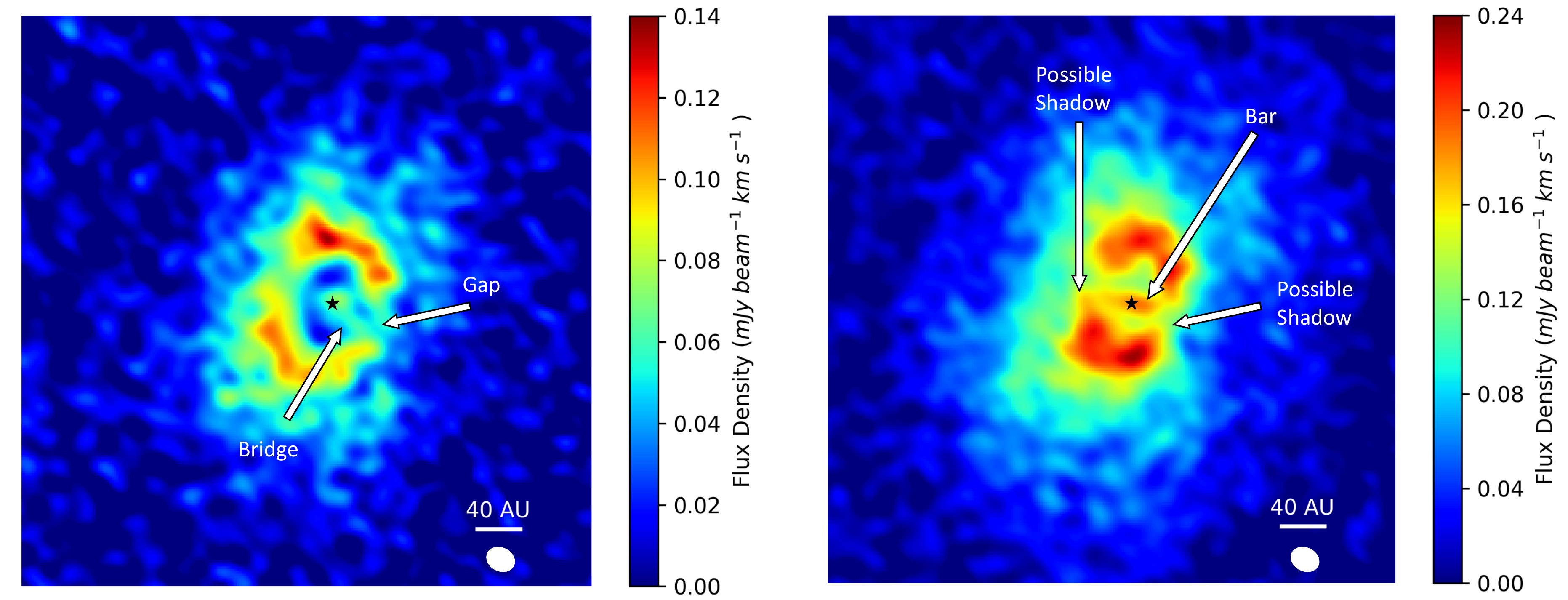}
\caption{Left: HCO\textsuperscript{+} J=4--3 moment 0 map created using velocity channels from 1.69-9.42 km s\textsuperscript{-1} and using emission detected above approximately 3$\sigma$.  The arrows in the moment 0 data point out important features referenced in the text. The HCO\textsuperscript{+} J=4--3 central blob appears to be slightly offset from the center. Right: CO J=3--2 moment 0 map created using velocity channels from 0.06-12.76 km s\textsuperscript{-1} and using emission detected above approximately 3$\sigma$. There appears to be a bar going across the center of the disk and what may be shadows at each end of the bar. These data are clipped and we have selected only the emission channels to generate these maps. The black star is at the location of PDS 70.}
\label{fig:gas_data0}
\end{figure}

The HCO\textsuperscript{+} J=4--3 moment 0 emission (Figure~\ref{fig:gas_data0}) reveals a similar gap to what is seen in the dust continuum (Figure~\ref{fig:continuum}), though it extends to only $\sim33$ au while the outer edge of the HCO\textsuperscript{+} J=4--3 moment 0 emission extends to $\sim160$ au. The gas disk also extends out farther than the dust though not as far as in many very young pre-main-sequence (PMS) stars such as DM Tau \citep{oberg10}. It contains a bright blob ($\sim$ 10 $\sigma$) to the N, as well as a bridge structure connecting the inner region and the outer disk in the SW, and a gap due W which lies directly adjacent to the bridge. There is also evidence of an azimuthal gap or depleted region to the NE. 

\begin{figure}[ht]
\centering
\includegraphics[scale=0.4]{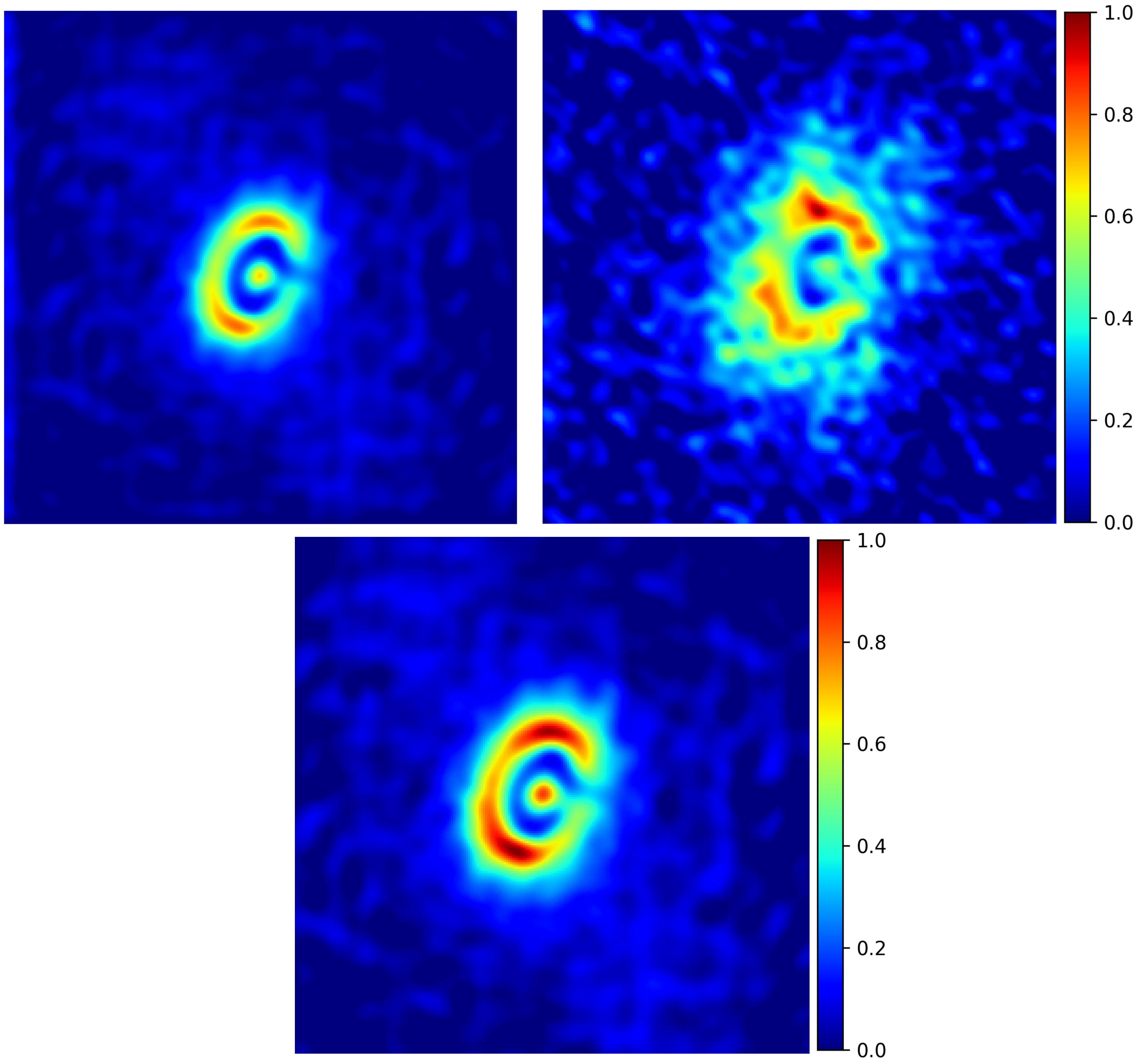}
\caption{Left: Simulated model image produced using the simobserve task in CASA 5.2. The image is displayed on the same scale as the right image with the peak brightness adjusted to match the brightness of the southern section of the right image. This was done to better show the relative brightness of the ring and central region because the bright blob in the north of the right image makes the comparison unclear. Right: HCO\textsuperscript{+} J=4--3 moment 0 map normalized to the brightest pixel in the image. Bottom: The same image as the top left, but normalized to the brightest pixel in the the image.}
\label{fig:Simulation}
\end{figure}

The bridge structure is likely to be real, with a signal-to-noise ratio of $\sim$ 5. In order to help verify this we generated a rough model of the bridge and azimuthal gap (Figure~\ref{fig:Simulation}). This was produced using the CASA 5.2 task simobserve with the thermal noise set to default and the antennas set to the ALMA cycle 4 configuration, alma.cycle4.3.cfg. The seed image for the model is a simple, two-component system generated using HOCHUNK3D with the bridge and depleted region added to the seed using image editing software. We found that in order to produce any asymmetric features to the degree of those seen in Figure~\ref{fig:gas_data0}, the inclusion of dimmer region to reproduce the gap and a bridge connecting the inner and outer disks was needed.

\begin{figure}[ht]
\centering
\includegraphics[scale=0.4]{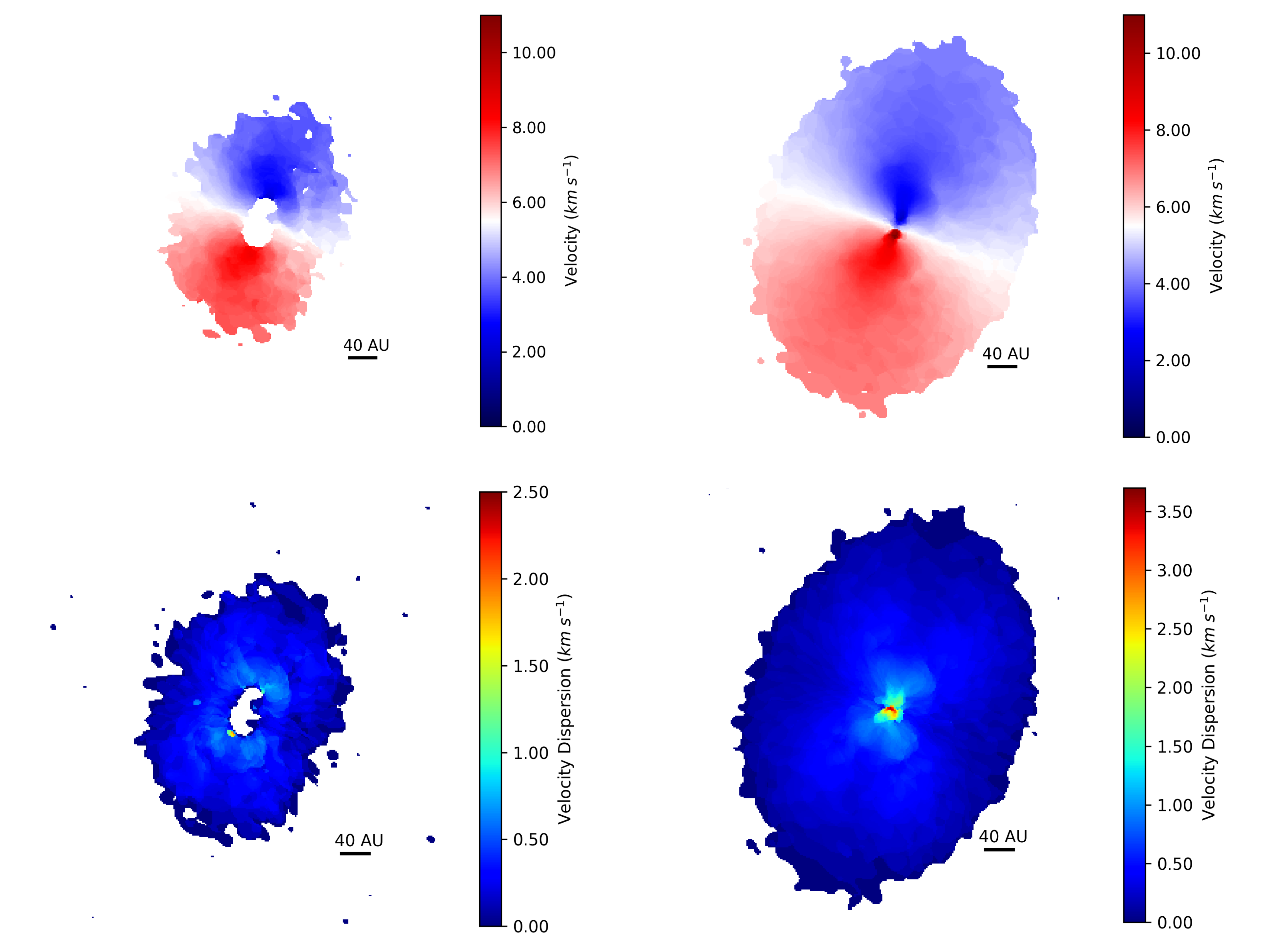}
\caption{Left top: HCO\textsuperscript{+} J=4--3 moment 1 map created using velocity channels from 1.69-9.42 km s\textsuperscript{-1} and emission above 0.028 Jy beam\textsuperscript{-1}. Left bottom: HCO\textsuperscript{+} J=4--3 moment 2 map created using velocity channels from 1.69-9.42 km s\textsuperscript{-1} and emission above 0.022 Jy beam\textsuperscript{-1}. Right top: CO J=3--2 moment 1 map created using velocity channels from 0.06-12.76 km s\textsuperscript{-1} and emission above 0.023 Jy beam\textsuperscript{-1}. Right bottom: CO J=3--2 moment 2 map created using velocity channels from 0.06-12.76 km s\textsuperscript{-1} and emission above 0.02 Jy beam\textsuperscript{-1}. All maps were created using CASA 5.1.}
\label{fig:gas_data12}
\end{figure}

The bridge feature itself appears to be mass transfer between the inner and outer disks similar to what is suggested for HD 142527 \citep{casassus13}. If this were the case we would expect to see a radial velocity component roughly perpendicular to the Keplerian motion of the disk which is not seen here (Figure~\ref{fig:gas_data12}). It has also been suggested that bridges such as this can be caused by a polar outflow of gas \citep{alves17}. Again however, this is likely not the case however as there appears to be no significant change in velocity in the region of the bridge, and it truncates at the outer disk instead of continuing to expel matter outside of the disk. Its close proximity to the western gap suggests that the two are related and could be due to large body formation which has been shown to cause both depleted regions as well as spiral arms in other disks \citep{edgar08, dong15, matter16}. Interestingly enough however, there is no evidence of the bridge feature or an azimuthal gap in the 0.87 mm continuum ring or in $H$-band polarized imagery \citep{dong12}. 

\begin{figure}[ht]
\centering
\includegraphics[scale=0.7]{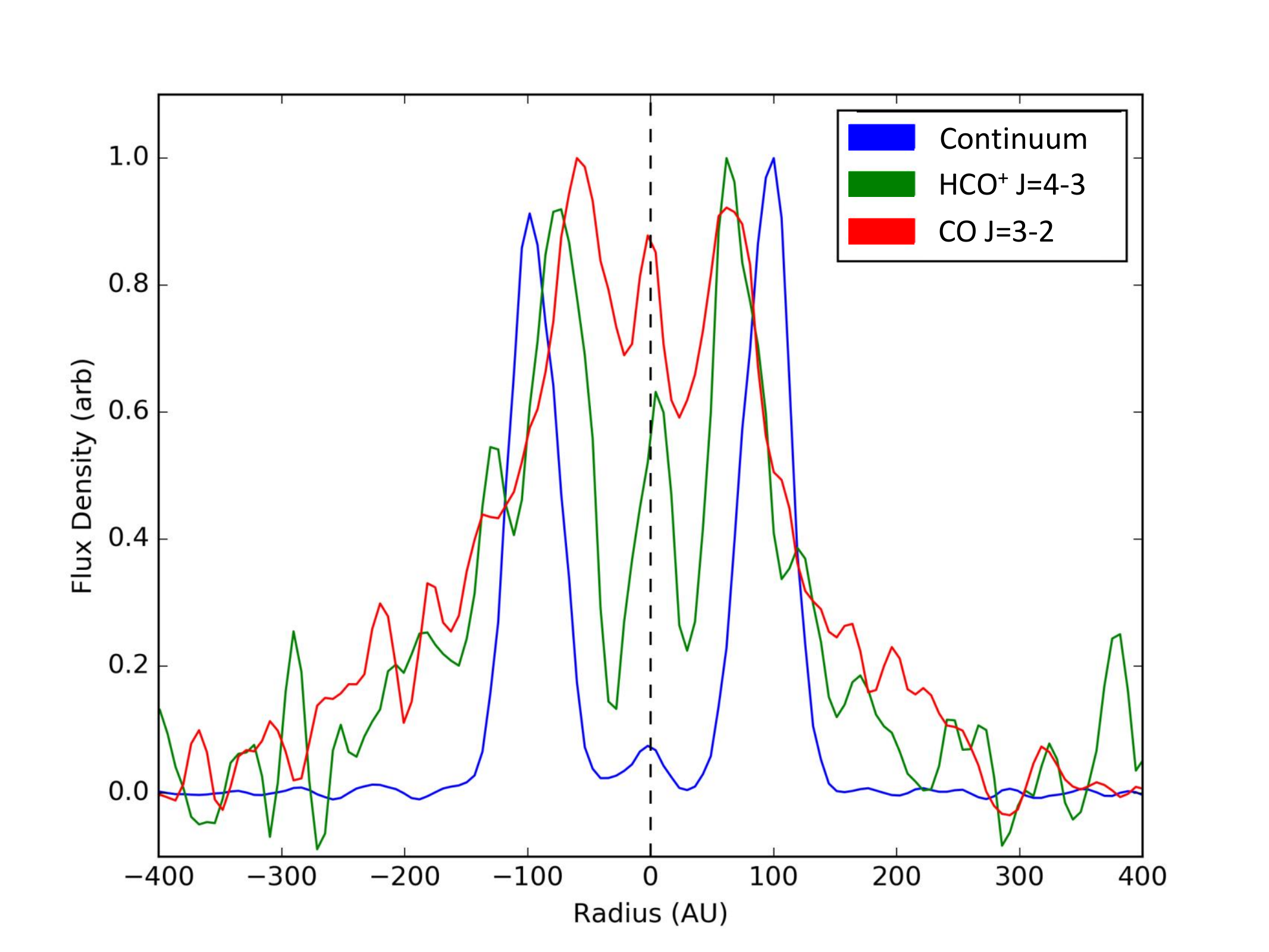}
\caption{Surface brightness slices along the major axis of the 0.87 mm continuum image (blue), HCO\textsuperscript{+} J=4--3 moment 0 data (green), and CO J=3--2 moment 0 data (red).}
\label{fig:sb_slices}
\end{figure}

\begin{figure}[ht]
\centering
\includegraphics[scale=0.6]{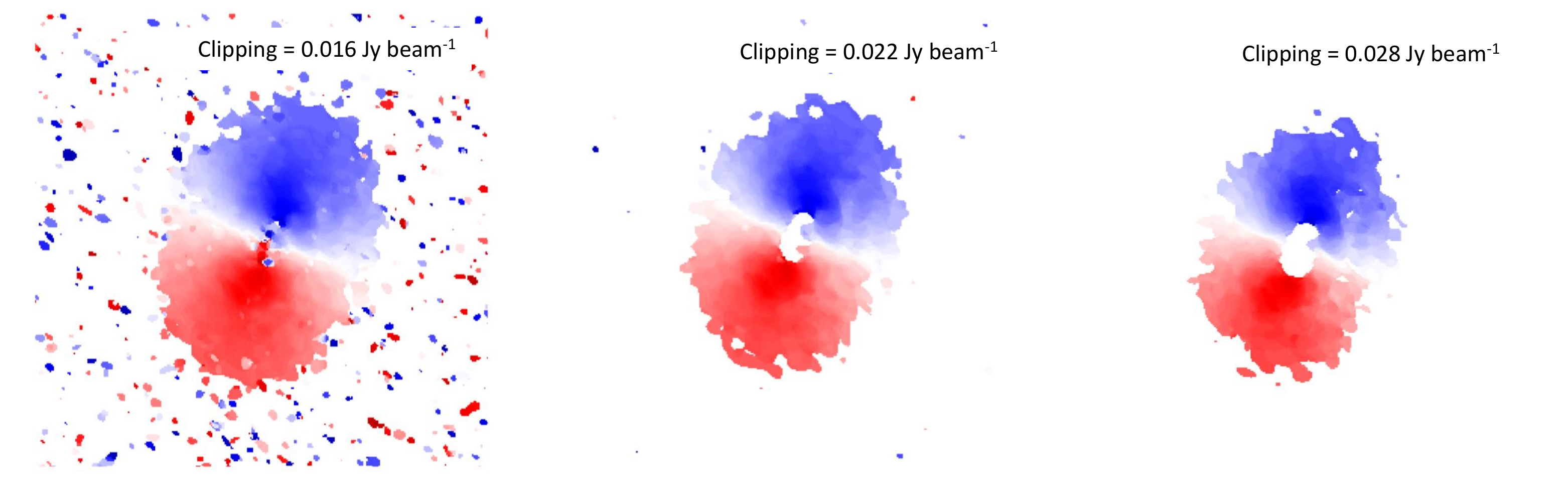}
\caption{HCO\textsuperscript{+} J=4--3 moment 1 maps at various clipping thresholds.}
\label{fig:clipping}
\end{figure}

As can be seen in Figure~\ref{fig:gas_data0}, the inner HCO\textsuperscript{+} J=4--3 disk appears slightly offset from the inner dust disk. To verify this we created radial traces of the continuum and moment 0 gas data, shown in Figure~\ref{fig:sb_slices}. There also may be some asymmetry in the outer gas disk as the HCO\textsuperscript{+} J=4--3 moment 0 peak intensities lie at different distances from the center of the disk. In the HCO\textsuperscript{+} J=4--3 moment 1 and 2 maps, the central region was clipped out due to excess noise in the region (Figure~\ref{fig:clipping}). 

\subsubsection{CO J=3--2 Measurements}

The CO J=3--2 moment 0 emission lies interior to the continuum and HCO\textsuperscript{+} J=4--3 emission, exhibiting little to no gap between the inner and outer disk. Similar to the HCO\textsuperscript{+} J=4--3 moment 0 data this disk extends further out than the dust disk, to $\sim210$ au, but again not as much as much younger PMS stars. This is promising for planet formation similar to what is seen in our solar system as there is still mass within 40 au of the star. The emission contains a blob in the S similar to what is seen in HCO\textsuperscript{+} J=4--3, but almost directly opposite its location in the disk. 

Again, similar to what is seen in HCO\textsuperscript{+} J=4--3, there is a bar feature towards the center of the image. This, however, appears to extend to either side of the disk and, as CO J=3--2 is generally considered optically thick in protoplanetary disks, could be causing the dimmed regions seen at either end of the bridge. It appears the bridge feature lies within the plane of the disk as the moment 1 data shows no deviation from Keplerian rotation in the region of the bridge (Figure~\ref{fig:gas_data12}).

The CO moment 2 map contains a butterfly pattern surrounding the central region. This pattern exists in the same locations as the four bright regions in the moment 0 data, but are due to crowded isovelocity curves in the moment 1 data and are likely not related.

\subsection{Model}

\subsubsection{Initial Model}
For our modeling we used the MCRT code HOCHUNK3D \citep{whitney13}. This code allows for two distinct grain populations as well as an envelope which can be manipulated into a third distinct grain population if necessary. The locations of these populations can also be changed to produce radial gaps in the disk.

We used a model similar to those of \citet{hashimoto15} and \citet{dong12} as our starting point. For the star we used a pre-main-sequence star with spectral type K5, a radius of 1.39 R$\astrosun$, and a temperature of 4400 K as suggested by \citet{gregorio02}. We also use a mass of 0.82 M$\astrosun$, estimated by \citet{riaud06} using spectral type, and further supported by dynamical fitting \citep{hashimoto15}. We then constructed a disk consisting of two rings with a large gap in the middle, a mass of 4.5 M\textsubscript{J}, an inclination of 50$\degr$ \citep{hashimoto12}, and a large grain to total grain mass ratio of 0.9667 in order to decrease the optical thickness of the small grains and increase the 10 and 20 $\mu$m silicate peaks. We also used the same grain types as the \citet{hashimoto15} model for the large and small grains which make up the disk.

Our initial model consisted of one that included both grain types in the inner disk which extended from the sublimation radius $\sim$0.05 au to 1 au which was used to reproduce the near IR emission seen in the SED. This was followed by a gap which extended to 65 au for the smaller grains and 80 au for the larger grains based on the $H$-band data \citep{dong12}, the 1.3 mm data \citep{hashimoto15}, and a distance of 140 pc \citep{riaud06}. This reproduced a fit to both the SED and 1.3 mm image seen in \citet{hashimoto15} as well as \citet{dong12}. 

\subsubsection{Model Fitting} 

Since the population of small grains is constrained by the character of the silicate emission bands near the 10 and 20 $\mu m$ wavelengths, we also include data obtained with the $Spitzer$ InfraRed Spectrograph (IRS) in our analysis. Both of these features are the result of warm silicate emission close to the star and required modifying the initial model to provide a reasonable fit. 

It is important to note also that the AKARI 18 $\micron$ point disagrees with the $Spitzer$ spectrum despite the $13.9-23.6$ $\micron$ band pass of the filter, and could indicate that either there is some variability of the disk or one of the data sets is incorrect. The grains which produce this peak lie approximately 0.5-3 au from the star which, assuming that the star is 0.82 M$\astrosun$, suggests orbital periods $\sim$0.39$-$5.7 years. It is possible therefore, that the disk could have changed between the two data sets. Additionally, it has been suggested that thermal instabilities due to active accretion could result in variability \citep{sitko12}. However PDS 70 seems to exhibit no accretion (Appendix A).

A second difference is the inclusion of the newly-acquired of ALMA 0.87 mm continuum data described in Section 2. The ALMA data have a much higher resolution than the SMA data used by \citet{hashimoto15}, which allows for much better spatial estimates of the disk. These data exhibit a different, $\sim$60 au, gap radius and lack the highly asymmetric spatial distributions in the SMA dust emission. This is in agreement with the gap radius found in scattered light \citep{dong12} which suggests that the apparent difference in gap radius for the two grain populations is not real.

To fit the model with these new constraints we included a less dense, though still optically thick, inner disk disk comprised of only small grains which extends from the sublimation radius to 15 au. We also raised the disk mass from 4.5 M\textsubscript{J} to $\sim7.3$ M\textsubscript{J} to fit the ALMA 0.87 mm point and raised the large grains to total grains ratio to 0.995.  This produced very strong silicate peaks, but does not reach the depth of the dips before the 10 and 20 $\micron$ peaks. The inner disk is followed by a gap extending to 60 au and an outer disk extending from 60 au to 110 au, based on the 0.87 mm continuum image. The higher spatial resolution of the ALMA data also allows for better comparison of the disk inclination to the model. We found the apparent inclination of the disk is best reproduced with a model inclination of 45$\degr$ from face on.

\begin{figure}[ht]
\centering
\includegraphics[scale=0.7]{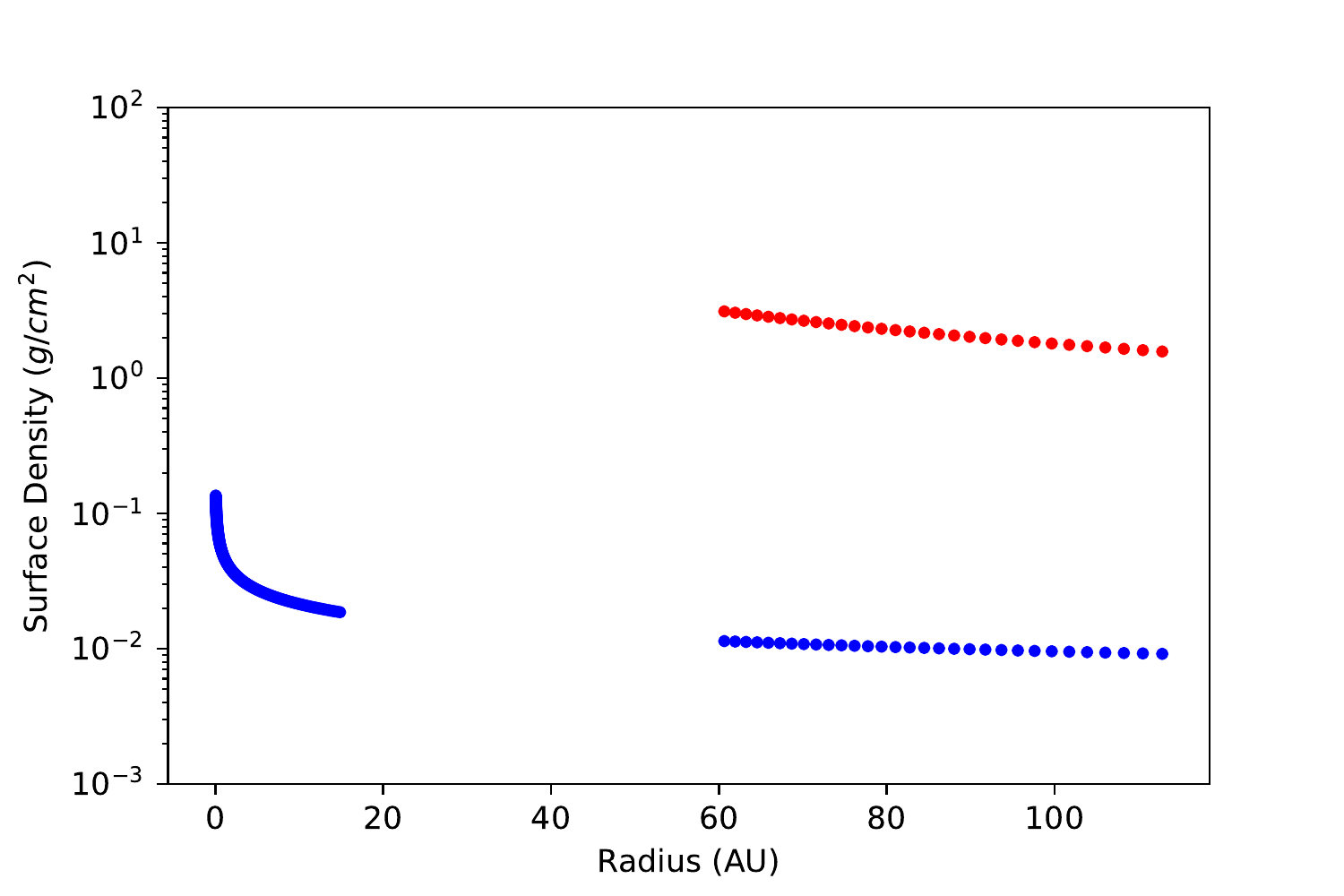}
\caption{Surface density profile for the outer disk of PDS 70. The blue dotted line is for the small grains disk and the red dotted is for the large grains.}
\label{fig:sigma}
\end{figure}

In order to probe the optical depth of the outer disk we examined the thickness of the disk, defined by $z = C_1 r^b$ where z is the scale height (thickness) of the disk, $C_1$ is a constant, r is the radial distance from the star, and b is the flaring exponent \citep{whitney03a}. Similarly the radial density is defined by $\rho = C_2 r^{-a}$ where $\rho$ is the density of the disk, $C_2$ is a constant, r is the radial distance from the star, and a is the density exponent \citep{whitney03a}. To match the SED we find b = 1.15 $\pm$ 0.02 and a = 2.25 $\pm$ 0.02 for the large grains in our disk and b = 1.15 $\pm$ 0.02 and a = 1.5 $\pm$ 0.02 for the small grains. Based on these parameters we find that the optical depth of the outer disk is $\tau \sim100$ au which is larger than the thickness of the outer disk suggesting that the disk is mostly optically thin at 0.87 mm. These two exponents can also be used to calculate the surface density of the disk via the method described in \citet{whitney13} eq. 15 ($\Sigma = \Sigma_0 \varpi^{-p}$) where $\Sigma_0$ is a constant set by the mass of the disk, $\varpi$ is the cylindrical radius, and $p=\alpha-\beta$. The surface density profile for the disk can be seen in Figure~\ref{fig:sigma}.

\begin{figure}[ht]
\centering
\includegraphics[scale=0.5]{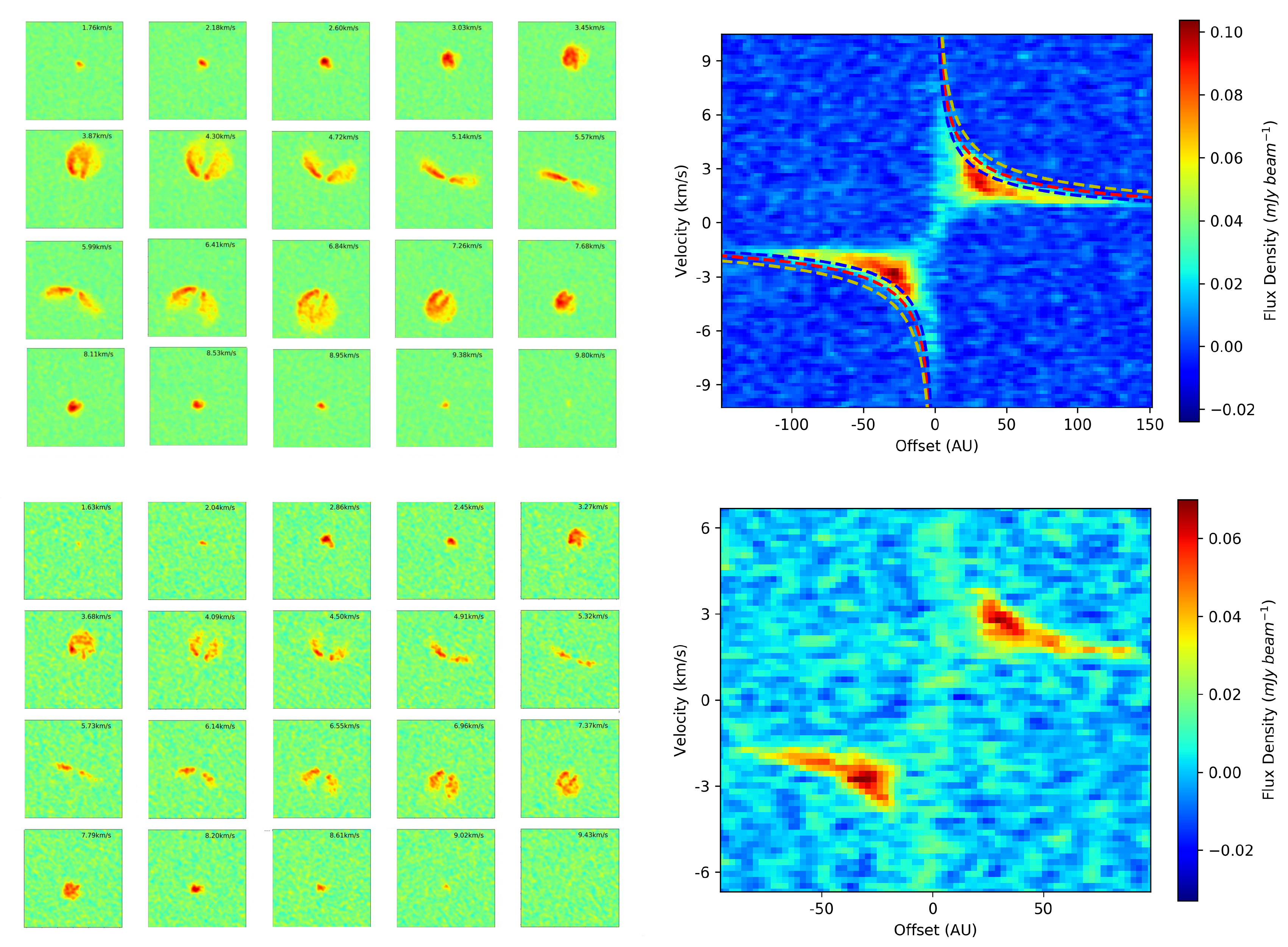}
\caption{Top Left: Channel map for CO J=3--2 emission. Top Right: Position velocity diagram of the CO J=3--2 moment 1 data. The blue dotted line is a Keplerian fit using an inclination of 45$\degr$ and a mass of 0.6 M$\astrosun$. The red dotted line is a Keplerian fit using an inclination of 45$\degr$ and a mass of 0.8 M$\astrosun$. The yellow dotted line is a Keplerian fit using an inclination of 45$\degr$ and a mass of 1.1 M$\astrosun$. Bottom Left: Channel map for HCO\textsuperscript{+} J=4--3 emission. Bottom Right: Position velocity diagram of the HCO\textsuperscript{+} J=4--3 moment 1 map.}
\label{fig:dynamical_mass}
\end{figure}

Because of the higher resolution of our data and the differences seen between this and the SMA data \citep{hashimoto15}, such as the outer disk extent, we used the CO J=3--2 moment 1 data to find the dynamical mass of PDS 70. Here we assumed a disk inclination of 45$\degr$, based on our modeling, and similar to what is found in \citet{hashimoto15} we see that a mass of 0.6 - 0.8 M$\astrosun$ (Figure~\ref{fig:dynamical_mass}) produces a good fit to the data and is in agreement with the \citet{riaud06} estimation. This falls within the range of 0.5 - 1.1 M$\astrosun$ found from comparing the location of PDS 70 to isochrones (Appendix) derived in \citet{tognelli11} and seen in Figure~\ref{fig:TTS_mass}.

\subsubsection{Final Model}

\begin{figure}[ht]
\centering
\includegraphics[scale=0.5]{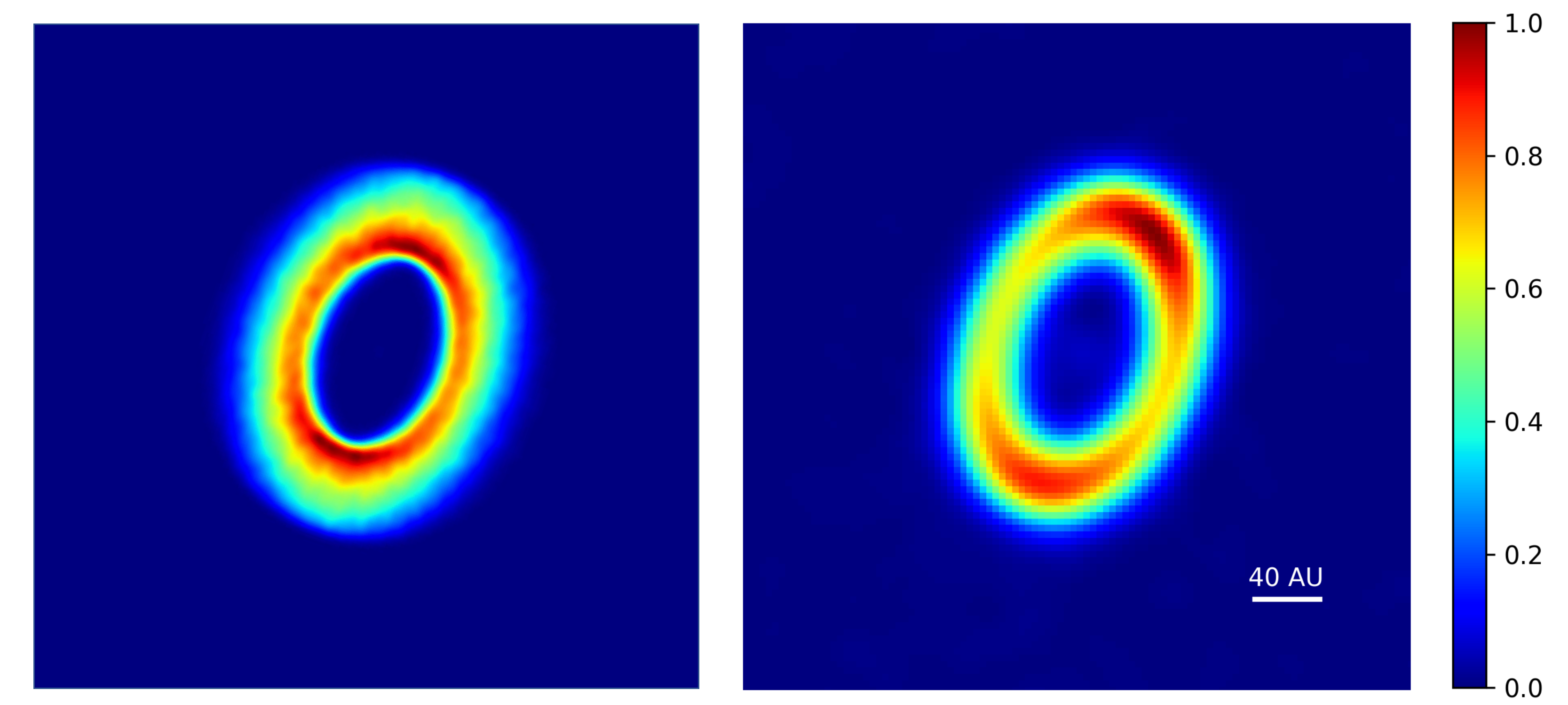}
\caption{Left: Model 0.87 mm image convolved with the ALMA beam shape. Right: ALMA 0.87 mm continuum image. North is up and East is left in these images. Both images are normalized to the brightest pixel in the image}
\label{fig:model_87}
\end{figure}

\begin{figure}[ht]
\centering
\includegraphics[scale=0.37]{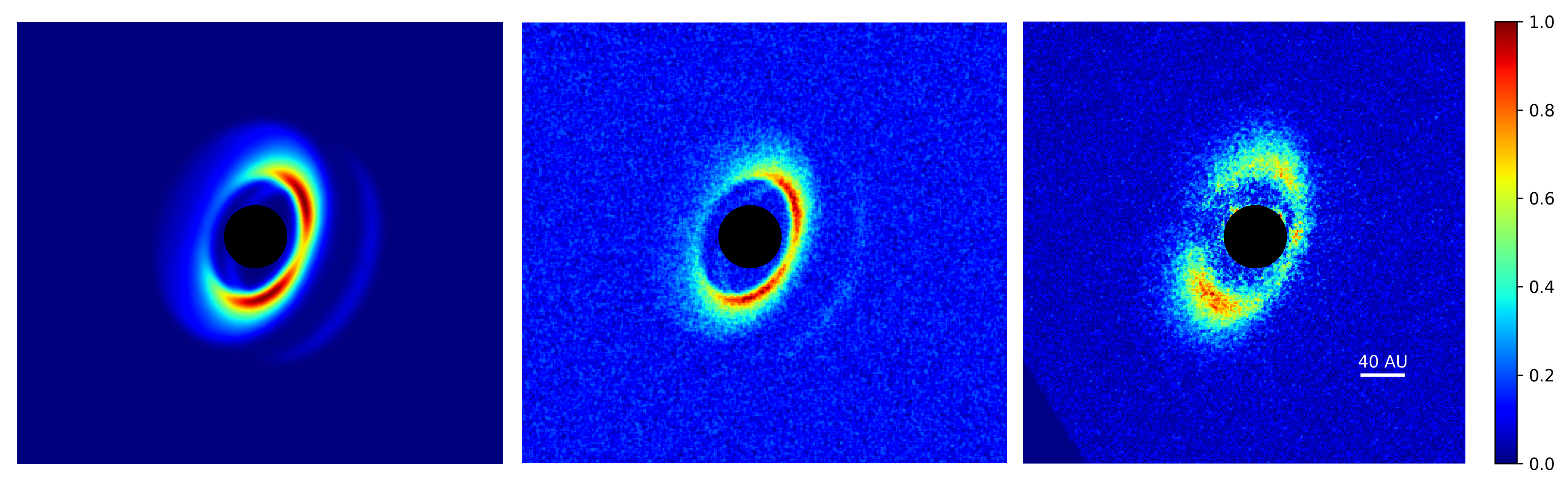}
\caption{Left: Model $H$-band polarized image convolved with the point spread function of the $H$-band image. Center: Same as left with additional background noise and photon noise added in to simulate what is seen in the HiCIAO image (right). Right: HiCIAO $H$-band polarized image. North is up and East is left in these images. The cavity wall is shadowed by the inner disk. Both images are normalized to the brightest pixel in the image. Differences in apparent radial extent between the model and data are likely due to noise in the data similar to what is seen in the center image.}
\label{fig:model_hband}
\end{figure}

To test the accuracy of our model, we compared model images to both the 0.87 mm continuum data (Figure~\ref{fig:model_87}) and the $H$-band data \citep{hashimoto12} (Figure~\ref{fig:model_hband}) after convolving with beam shape and point-spread function of these data respectively. In the case of the 0.87 mm model image we see a ratio of emission between the inner and outer disks which is in agreement with the ALMA image, \(\frac{Finner}{Fouter}\) $\sim0.004$ in both cases, which suggests the inner disk is essentially devoid of large grains. The $H$-band data reproduces the apparent dip in brightness seen in the NE side of the disk (Figure~\ref{fig:model_hband}). This dip is the result of co-planar shadowing from the the inner disk. The final model parameters can bee seen in Table~\ref{tbl:parameters}.

\section{Discussion}

\subsection{Asymmetry in 0.87 mm Continuum Image}
If we assume the 0.87 mm continuum is optically thin, section 3.2.2, the difference in brightness should correspond to a difference in mass density between the two locations, suggesting that there is more mass density in the NW side of the outer disk. There exists the possibility however, that the difference in brightness could be a result of apocenter glow \citep{pan16}, though we do not detect a pericenter offset in this system. Given the resolution of the image however, there could be an offset up to $\sim5$ au which would result in an apocenter glow of $\sim6.7\%$  according to the calculation proposed in \citet{pan16}.

\subsection{Differences in Gas and Dust Imagery}

Most of the gas features seen in the HCO\textsuperscript{+} J=4--3 and CO J=3--2 maps do not have an analogue in the 0.87 mm continuum or $H$-band imagery. The only notable exceptions are the excess flux in the NW of the continuum image, an inner gap, and an inner disk which is present in all available data.

\subsubsection{Azimuthal Gap in HCO\textsuperscript{+} J=4--3}

While some of these features are easily explained by known properties, such as the diminished, central region in the CO J=3--2 emission and the larger radial extent of both HCO\textsuperscript{+} J=4--3 and CO J=3--2 emission, of the gas the lack of an azimuthal gap in the continuum or $H$-band data is not. It is likely not a gravitational depletion of the disk by a compact body as there should be no difference in how the dust and gas are affected gravitationally. This suggests a more exotic cause for the apparent difference.

Options we have considered are a large stellar flare from a coronal mass ejection, an embedded planet, or shadowing from an inner optically thick blob of material. A large stellar flare could disturb the gas in a localized region of the disk and possibly move the dust in that region as well. This could neatly link the azimuthal gap seen in the HCO\textsuperscript{+} J=4--3 data as well as the large blob of mass seen in the 0.87 mm continuum.  A similar effect was seen in AU Mic where it is suggested that a stellar flare caused fast moving features in the dust of the disk \citep{boccaletti15}. However, though this could cause feature propagation in the dust disk, we would expect the density wave to travel around the disk or perhaps cause dust blowout from the impact. The difference in density from either scenario is not seen in the 0.87 mm continuum and there seems to be no difference in the location of the bright blob in the 1.3 mm continuum \citep{hashimoto15} and the 0.87 mm continuum.

An embedded planet could cause the local region around it to undergo various chemical changes due to heating from the planet \citep{cleeves15}. Though the heating would initially cause any frozen out gasses to sublimate, it may also catalyze complicated chemical interactions which could locally deplete the HCO resulting in an apparent gap in the gas disk. It has been shown that a Jupiter-sized planet can cause an effect in a radius of several au surrounding the planet \citep{cleeves15} which grows as the planet moves further from the star. Unfortunately a planet of this size would not be visible in the currently available data.

Shadowing from an inner disk would result in an apparent gap in the disk dust \citep{long17,stolker16} and should similarly effect the gas if the inner disk is optically thick enough. In this instance, however, the shadowing body would necessarily have an asymmetric distribution as there is no equivalent azimuthal gap on the western side of the HCO\textsuperscript{+} J=4--3 disk, only minor dimming. This could explain the apparent inner disk offset seen in the HCO\textsuperscript{+} J=4--3 moment zero data as it lies on the NE side of the star closely aligned with the apparent gap. If this is the source of the apparent gap, then we would expect the shadow to move as the shadowing body moves around the star. This should happen within several months given that the shadowing body would appear to be at a radius of $\sim5$ au. 

\subsubsection{Gas-poor inner disk region}

\begin{figure}[ht]
\centering
\includegraphics[scale=0.6]{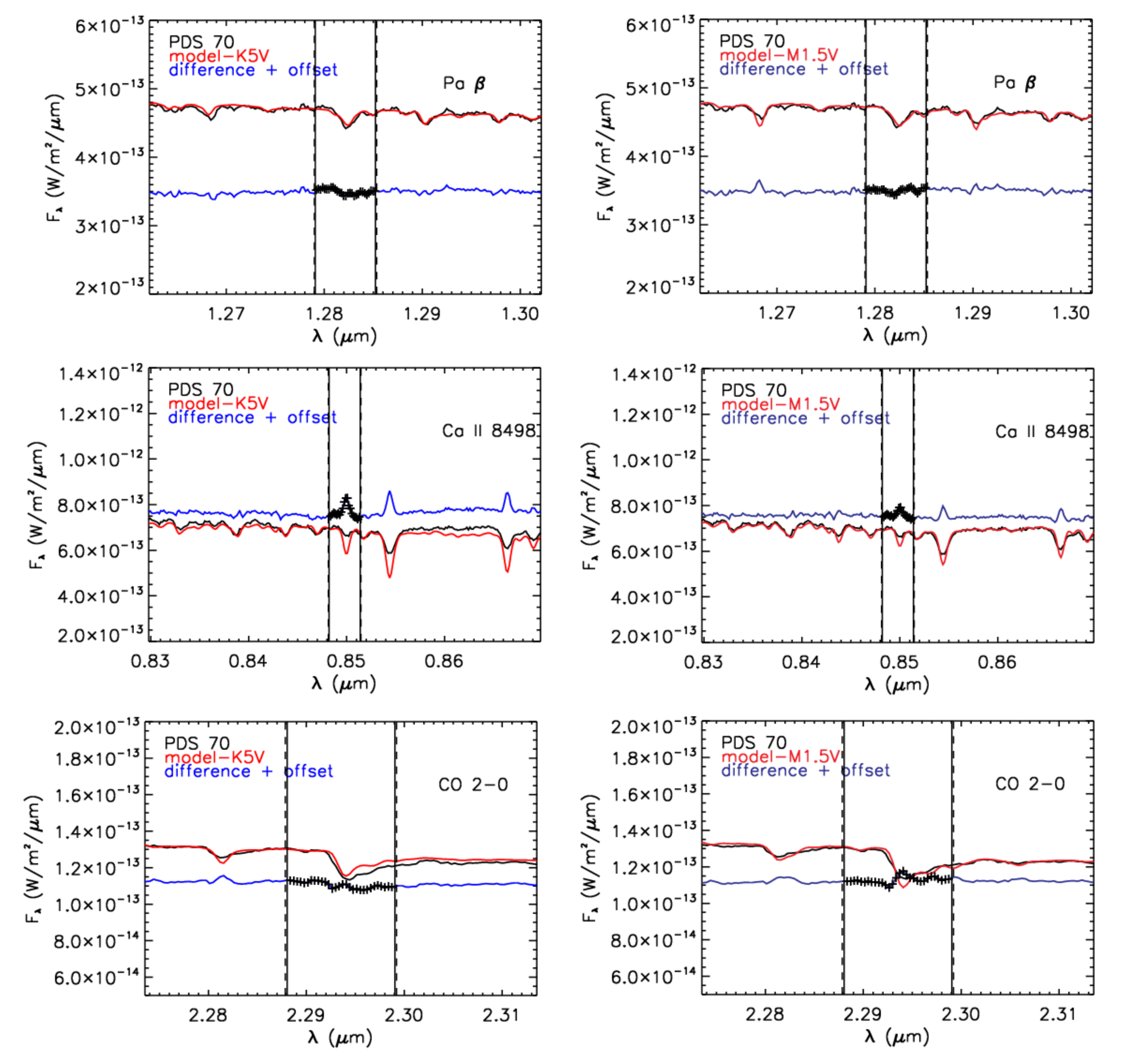}
\caption{Line extractions from SpeX for Pa $\beta$, CO (2-0), and Ca II 8498 for PDS 70, the red line is our best fit model for the emission, and the blue line is the difference between the two plus an offset. We performed line extractions for both K5V (left) and M1.5V (right) model stars.}
\label{fig:LineExtractions}
\end{figure}

As previously mentioned the SED suggests the existence of dust close to sublimation radius of the star which is further supported by our modeling. In contrast, we see in our SpeX data (discussed more thoroughly in the Appendix) that the innermost regions of the disk, less than 1 au, seem to be devoid of gas as there is no net emission for hydrogen or CO gas (Figure~\ref{fig:LineExtractions}) above photospheric. There is however some Ca II emission which suggests that there is chromoshperic activity similar to what is seen in HD 163296 \citep{sitko08}. The lack of gas emission close to the star suggests that there is also no active accretion of gas onto the star.



\subsubsection{Bar across CO}

The CO J=3--2 moment 1 data reveals what appears to be a bar along the minor axis of the disk.  At opposite sides of the disk, directly adjacent to the bridge feature there appear to be dips in brightness in the CO moment 1 map. If the bridge feature is optically thick, as expected from CO gas, it could in turn shadow the gas in the outer disk. This however raises the question as to how the bridge structure could have formed. It could be a mass transfer between the inner and outer disk, though over time we would expect that the bridge would not remain solely along the minor axis, but instead would smear out due to Keplerian motion. We would also expect that some dust would be transferred as well which is not seen in the continuum data. This suggests it could lie in a plane nearly perpendicular the plane of the outer disk. However, as mentioned in Section 3, there appears to be no deviation from Keplerian motion in the vicinity of the bridge. The lack of deviation also means it is likely not an outflow from the inner disk as well.

\section{Conclusion}

In this paper we presented the imagea of the dust continuum at 0.87 mm, the HCO\textsuperscript{+} J=4--3 emission, and the CO J=3--2 emission of the disks around PDS 70. We use these along with the SED of the system and $H$-band polarized data to produce a MCRT model of the system. The main results of our modeling and observation are as follows.

\begin{itemize}

\item There is an azimuthal gap in the HCO\textsuperscript{+} J=4--3 emission which has not been observed in other gas disks. Several possible causes include gravitational depletion by an embedded body, a large stellar flare, chemical changes due to the heating of an embedded body, and shadowing of the gas by an inner disk component. Further study is needed however to determine the most likely candidate.

\item The sub-au region of the disk is gas poor despite evidence for a dusty inner disk component. This suggests that there is likely no gas accretion onto the star.

\item There appears to be an arm in the HCO\textsuperscript{+} J=4--3 emission. The arm could be caused by a planet either very close to the star ($\sim$5 au), or in the radial gap seen in the HCO\textsuperscript{+} J=4--3 and the ring seen the continuum image.

\item The CO J=3--2 data exhibits a bridge along it's minor axis. The bars linear shape and the lack of deviation from Keplerian motion in the  moment 1 data in that region indicate that it lies within the plane of the disk, but has not been smeared out due to Keplerian rotation. The optical thickness of CO gas, and the close proximity of the bridge to the two dim regions in the CO J=3--2 emission, suggest the bridge could be shadowing the gas disk.

\item MCRT modeling, 0.87 mm continuum, and $H$-band polarized data suggest that PDS 70 has a two component dust disk. The inner disk component extends from the sublimation radius out to $\sim$15 au and is followed by a radial gap to $\sim$60 au. The outer disk component then extends from $\sim$60 au to $\sim$110 au. 

\item We do not observe a difference between the small and large dust grain locations in the outer disk as proposed in previous work. Our 0.87 mm continuum image, which probes large grains on the order of 1 mm in size, exhibits a similar radial extent to that of the $H$-band polarized imagery, which probes small grains. The difference between our findings and that of \citet{hashimoto15} is likely due to the low angular resolution of SMA data when compared to ALMA data.

\item The CO J=3--2 moment 1 measurements suggests a dynamical mass of PDS 70 of 0.6 - 0.8 M$\astrosun$ similar to what has been seen in \citet{hashimoto15} and is consistent with the extended kinematics around a K5 star. 
   
\end{itemize}

In this paper we have discussed various features which appear in the gas disk of PDS 70 but are not present in the dust disk. Although we cannot determine a definitive cause for these features with existing data, their existence does provide new incentives to observe this disk with other high angular resolution, high contrast instruments such as VLTI MATISSE or the NIRISS aperture masking interferometry mode of the James Webb Space Telescope. These instruments would allow us to probe the inner regions of the disk where we suspect large planets could be causing many of these features.

We also discussed the overall structure of the dust disk found from continuum data and MCRT modeling. The disk seems to have no strong segregation between the dust grain sizes in the outer disk, but the inner disk does appear to be devoid of large grains. This is contrary to what is suggested in previous work, but is supported by the large silicate peaks in the SED. Further investigation of the inner disk is necessary to confirm its radial extent and optical depth, and may be accomplished through use of the instruments mentioned above.

\acknowledgements

This work is supported supported by the NASA XRP grants NNX17AF88G and NNX16AJ75G. MC thanks the support from the Centro de Astrof\'isica de Valpara\'iso. S.K. acknowledges support from an STFC Rutherford Fellowship (ST/J004030/1) and ERC Starting Grant (Grant Agreement No.\ 639889). This work is supported by the Astrobiology Center Program of National Institutes of Natural Sciences (NINS) (Grant Number: AB281013) and by MEXT KAKENHI No. 17K05399 (E.A.). Y.H. is currently supported by Jet Propulsion Laboratory, California Institute of Technology, under a contract from NASA.

\bibliography{Long_Paper_PDS70_160318ver11}

\section{Appendix}

\subsection{Appendix A}

\subsubsection{SpeX and REM Observations of PDS 70}

PDS 70 was observed on 13 Jan. 2017 UT using the SpeX spectrograph \citep{rayner03} on NASA's Infrared Telescope Facility (IRTF). For these observations the SXD grating (0.7-2.4 $\mu$m) and 0.8 arcsec wide slit were used. Data reduction was done using the Spextool package \citep{cushing04} running under IDL. Telluric correction and flux calibration \citep{vacca03} were done using the A0V star HD 130163. 

\begin{figure}[ht]
\centering
\includegraphics[scale=0.6]{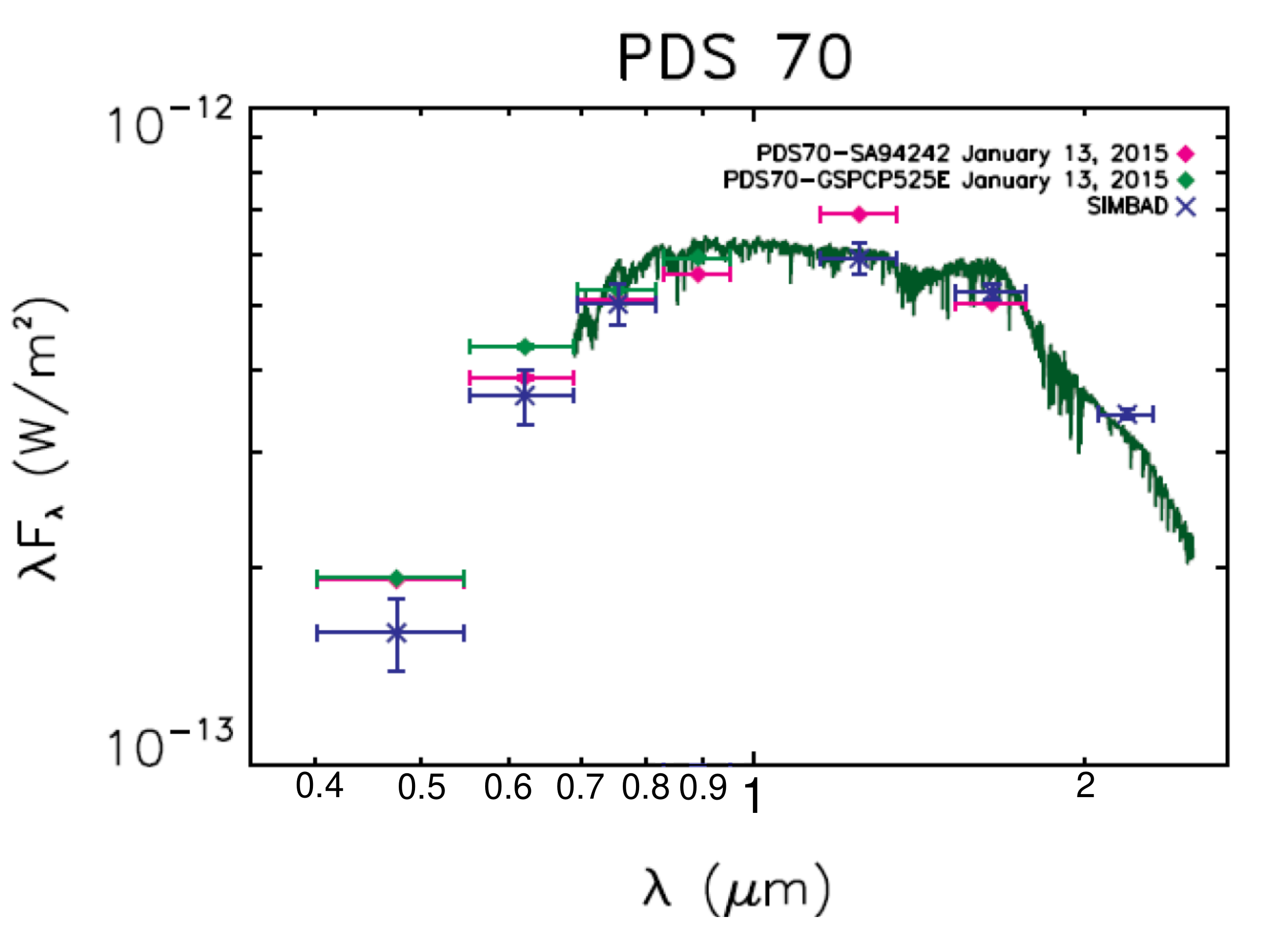}
\caption{REM Data in the g$\arcmin$r$\arcmin$i$\arcmin$z$\arcmin$JH filters, the green diamonds are data using the flux calibration star GSPC-P525E and the pink diamonds are data using the flux calibration star SA94-242. The blue Xs are both the  g$\arcmin$r$\arcmin$i$\arcmin$ data from the AAVSO Photometric All Sky Survey and JHK data from the 2MASS survey. The green line is the SpeX data for PDS 70.}
\label{fig:REM}
\end{figure}

We also observed PDS 70 on 10 June 2016 UT using the Rapid Eye Mount (REM) telescope in the g$'$r$'$i$'$z$'$JH filters. Two flux calibration stars were used, SA94-242 and GSPC-P525E. Our REM data is shown in Figure~\ref{fig:REM}, along with the g$'$r$'$i$'$ data from the AAVSO Photometric All Sky Survey (APASS; \citet{henden16}) and JHK data from the 2MASS survey \citep{skrutskie06}. These were used to determine the approximate absolute fluxes for the SXD data. 

\begin{figure}[ht]
\centering
\includegraphics[scale=0.7]{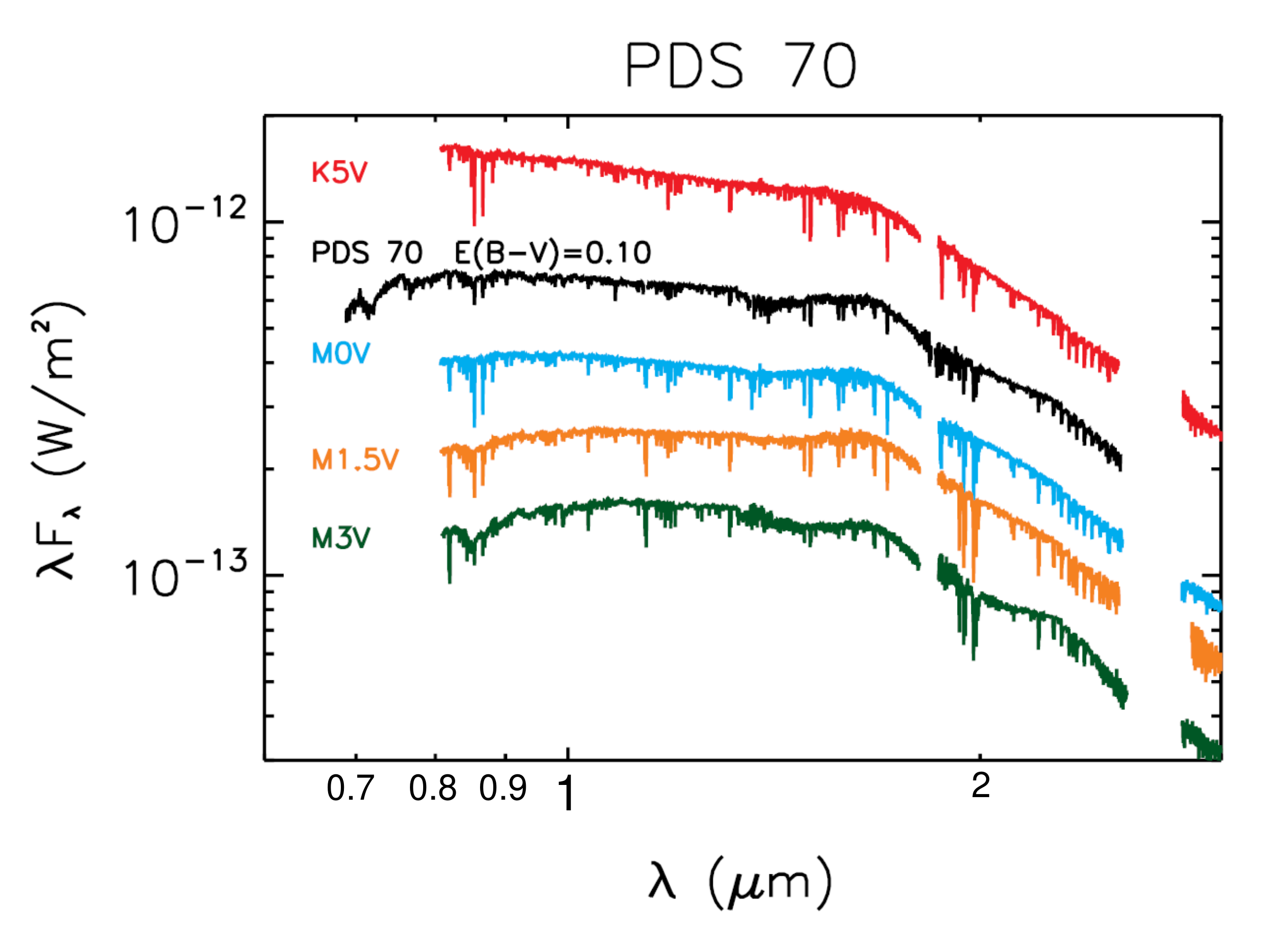}
\caption{Our SpeX spectrum data in black (dereddened by E(B-V)=0.10) with other spectrum data gathered from the IRTF Spectral Library for K5V, M0V, M1.5V, and M3V stars.}
\label{fig:Mstar}
\end{figure}

In Figure~\ref{fig:Mstar} we compare our SpeX spectrum with a sample of similar data sets from the IRTF Spectral Library (\url{http://irtfweb.ifa.hawaii.edu/~spex/IRTF_Spectral_Library/}). Despite K5V classification usually adopted for the star \citep{gregorio02,riaud06}, at near-IR wavelengths it resembles an early M star. Similar differences in classification between visible and infrared wavelengths have been seen before in other stars such as TW Hydrae \citep{debes13} and V4046 Sgr \citep{kastner14}. 

\subsubsection{Gas Accretion in the PDS 70 System}

\begin{figure}[ht]
\centering
\includegraphics[scale=0.5]{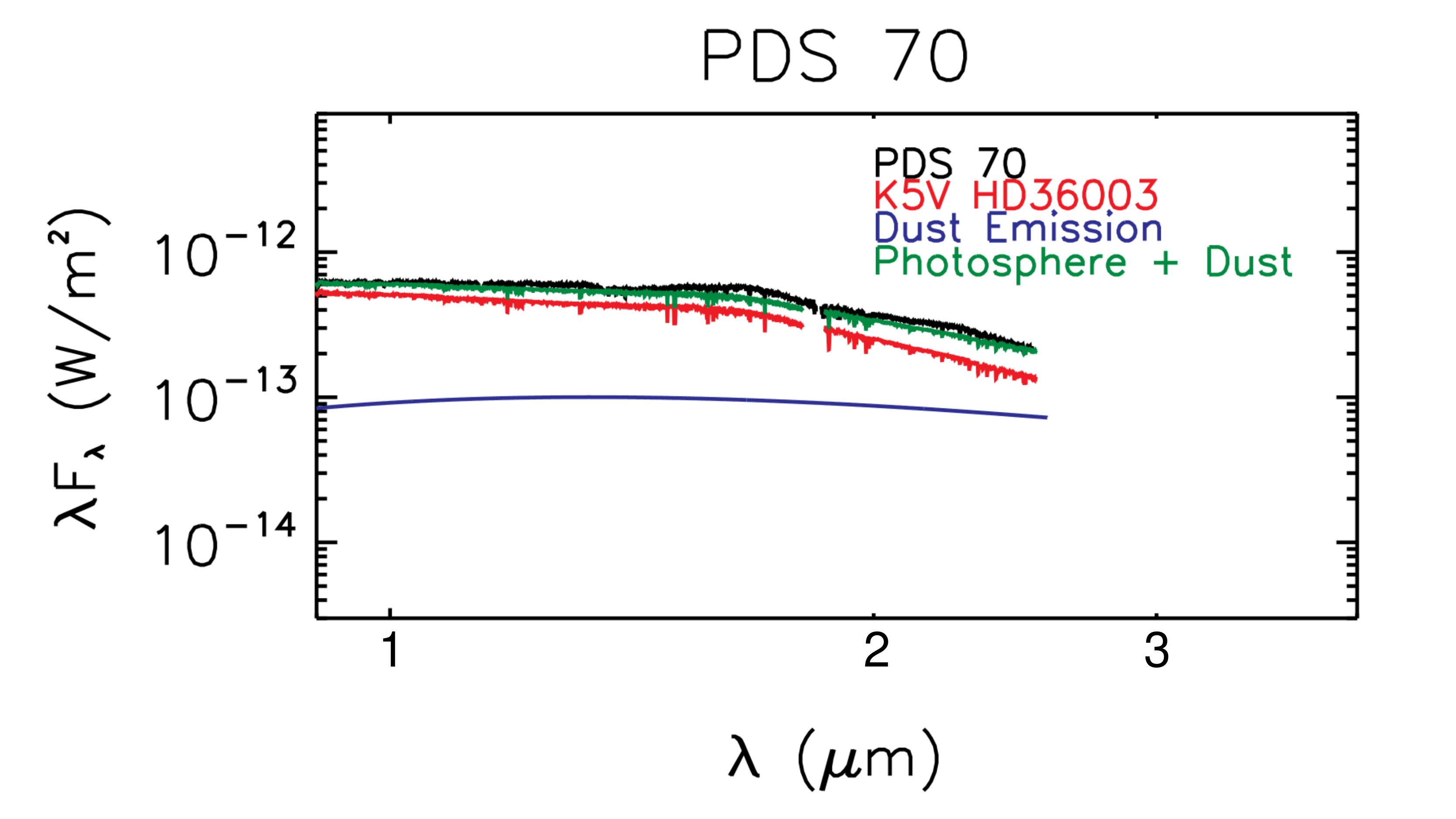}
\caption{Model for PDS 70. The SED consists of the photospheric emission of the star plus the  thermal emission by the inner dust. The relative strengths of these two components were adjusted to match the observed emission of the star plus disk system. The process used is fully described in \citet{sitko12}.}
\label{fig:Korash_SED}
\end{figure}

To determine the potential gas accretion characteristics in PDS 70, we subtracted off a spectral model consisting of the photospheric flux and the thermal contribution of the dust (Figure~\ref{fig:Korash_SED}), in a manner similar to that of \citet{sitko12}. For the template photosphere, we used the SpeX spectrum of a K5V star. Despite the fact that \citet{gregorio02} found H $\alpha$ emission, we find that the Pa $\beta$ line commonly used as a measure of accretion in PMS stars showed no detectable emission (Figure~\ref{fig:LineExtractions}). The CO bandhead emission between 2.3 and 2.4 $\mu$m that are sometimes seen in emission in gas-rich debris disks also look as if there is no appreciable CO in emission. In contrast, the lines of Ca II are weakly in emission. It is known that in late type PMS stars, lines of Ca II can be in emission when accretion is otherwise weak or absent \citep{ingleby11}, suggesting that this line, and possibly H $\alpha$, are chomospheric in origin, and not due to gas accretion onto the star. These data suggest that the inner regions of the disk system in PDS 70 are gas-poor, and that the CO gas detected with ALMA in the inner regions of PDS 70 do not extend down to the star itself. This is also suggested from CO J=3--2 moment 1 data where we can infer that the gas only extends down to $\sim5$ au.

\subsubsection{Mass and Age of PDS 70}

\begin{figure}[ht]
\centering
\includegraphics[scale=0.7]{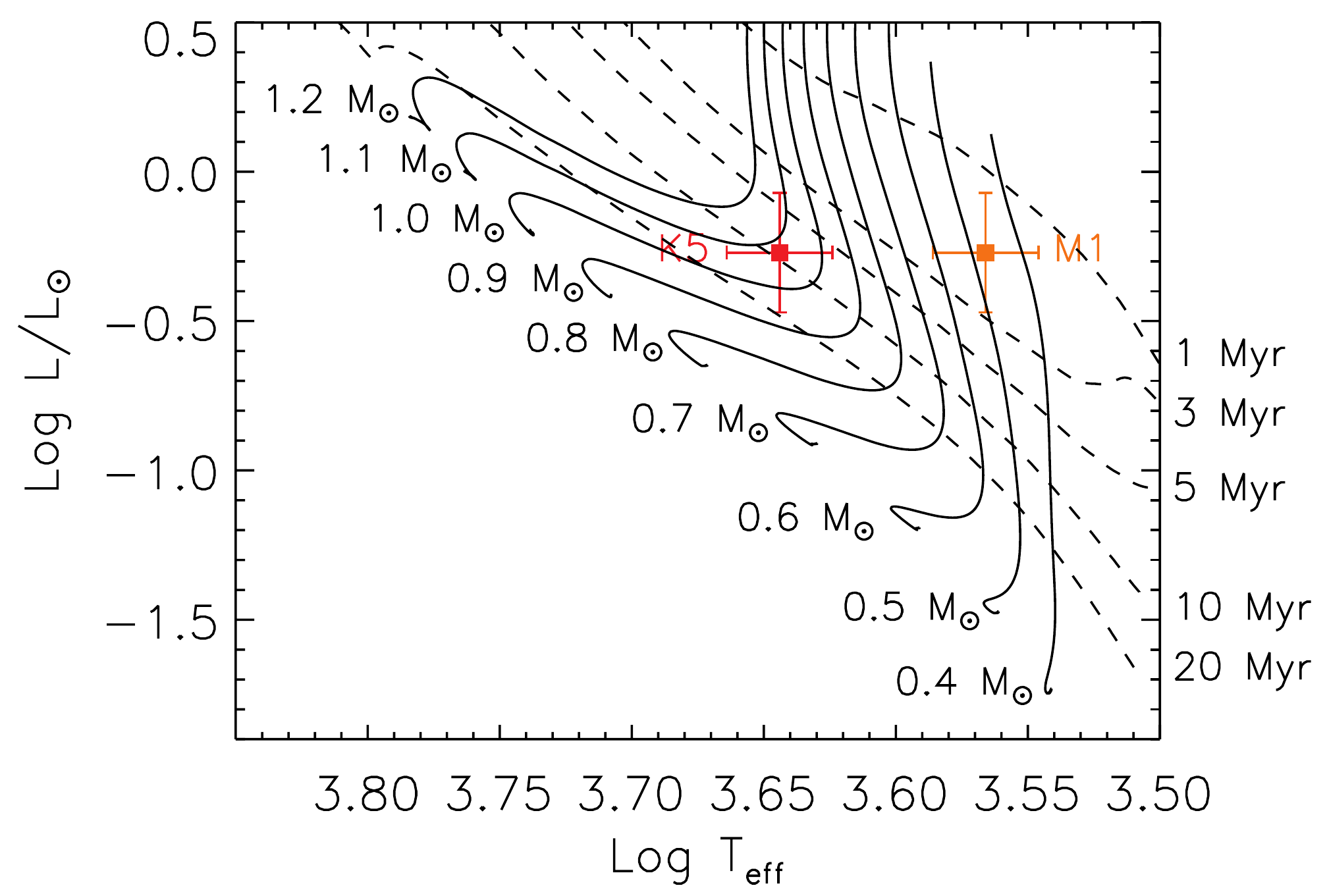}
\caption{The location of PDS 70 relative to the PMS evolutionary tracks and isochrones of \citet{tognelli11}. The two locations of PDS 70 are derived using an assumed spectral type of K5 (red point) and M1 (orange point), and produce different mass and age estimates.}
\label{fig:TTS_mass}
\end{figure}

Using the V magnitude of PDS 70 \citep{kiraga12}, the 140 pc distance from the association with the Sco-Cen Association,  and the bolometric correction and initially adopting an effective temperature for a K5 PMS stars from \citet{pecaut13} we derived a luminosity of 0.65 L$_{\astrosun}$. Figure~\ref{fig:TTS_mass} shows the location of PDS 70 relative to the PMS evolution tracks and isochrones \citep{tognelli11}, for z = 0.02, helium mass fraction of 0.27, convective parameter $\alpha$ = 1.68, and a deuterium abundance of $2e^{-5}$. The on-line versions of these use the abundances of \citet{asplund05} but are nearly indistinguishable from those based on the revised abundances of \citet{asplund09} as seen in the figures in \citet{tognelli11}. From these we derived a mass of 1.1 M$_{\astrosun}$ and an age of 10 Myr. If instead, we do the same determination for an M1V classification, the mass drops to 0.5 M$_{\astrosun}$ and the age to around 2 Myr. Based on the equivalent width of the Li lines in PDS 70 from \citet{gregorio02}, \citet{metchev04} derive an age of $\le$ 10 Myr. The location in the sky would put PDS 70 in the Upper Centaurus Lupus (UCL) region \citep{preibisch08} of the Sco-Cen Association. \citet{song12} have looked at a large number of G, K, and M stars in the Sco-Cen complex. Based on PMS evolutionary tracks and Li line strengths, they find an age of $\sim$ 10 Myr for this region. This number is consistent with the age we obtained for the K5V classification for PDS 70.

\begin{figure}[ht]
\centering
\includegraphics[scale=0.8]{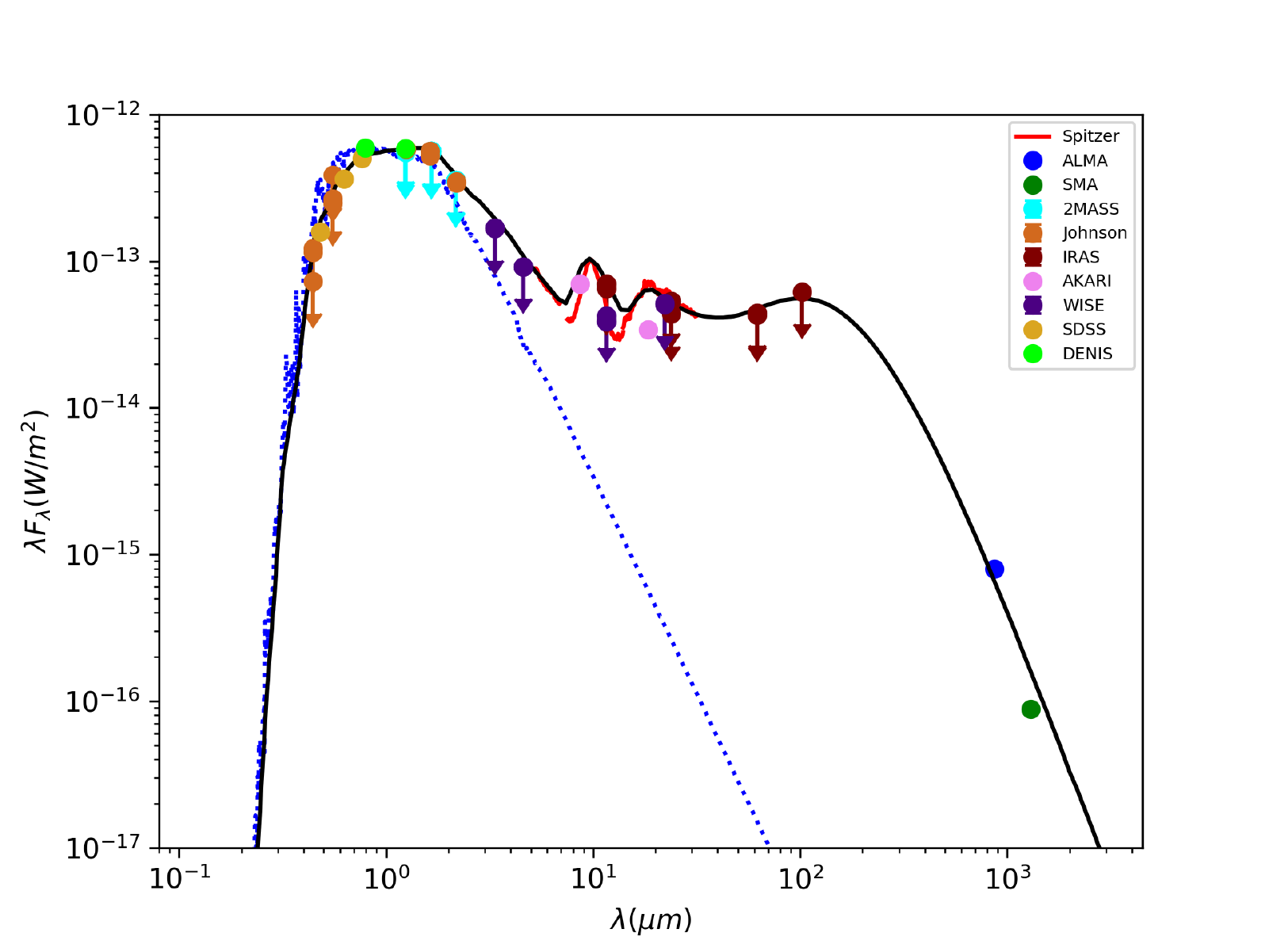}
\caption{SED of PDS 70. The blue dotted line is the photosphere of a 4400K star and the black line is the model fit.}
\label{fig:SED}
\end{figure}

\begin{table}
\begin{center}
\begin{tabular}{ |p{4cm}||p{4cm}|p{6cm}|  }
 \hline
 \multicolumn{3}{|c|}{Model Parameters} \\
 \hline
 Parameter & Value & Source\\
 \hline
 Stellar Radius & 1.39 R$\astrosun$ & \citep{gregorio02} \\
 Stellar Mass & 0.82 M$\astrosun$ & \citep{riaud06} \\
 Stellar Distance & 140 pc & \citep{riaud06}\\
 Stellar Temperature & 4400 K  & \citep{gregorio02}\\
 Spectral Type & K5V & \citep{gregorio02} \\
 Small Grains & ISM & \citep{kim94} \\
 SG Inner Radius & 0.05 au & -\\
 SG Outer Radius & 110 au & -\\
 Large Grains & Model 2 & \citep{wood02} \\
 LG Inner Radius & 60 au & -\\
 LG Outer Radius & 110 au & -\\
 Disk Inclination & 45  & -   \\
 Gap & 15-60 au & - \\
 Disk Mass & 0.42 M\textsubscript{J} & - \\
 \hline
\end{tabular}
\end{center}
\caption{The parameters used for the final model are shown here. For the small grains we used the standard ISM dust model from \citet{kim94} composed of silicates and graphite, a radial power-law exponent of $\sim$3.5, and a maximum particle size $\leq$ 0.2 $\mu$m. For the large grains we used Model 2 from \citet{wood02} composed of astronomical silicates and amorphous graphite, a radial power-law exponent of $\sim$3.5, and a maximum particle size $\leq$ 1 mm.}
 \label{tbl:parameters}
\end{table}

\end{document}